\documentclass{pazh}

\usepackage{graphicx}
\usepackage{latexsym}

\newcommand{\km}{\,\mbox{km}\,\mbox{s}^{-1}}
\def\Ha{\hbox{H$_\alpha$\,}}
\def\farcm{\hbox{$.\mkern-4mu^\prime$}}
\def\farcs{\hbox{$.\!\!^{\prime\prime}$}}
\def\SCORPIO{\mbox{SCORPIO}\,}

\begin{document}

\title{Structure and Kinematics of the Interstellar Medium in the Star-Forming Region
in the BCD~Galaxy VII~Zw~403 (UGC~6456)}

\author{T.A. Lozinskaya$^{1}$ \and A.V. Moiseev$^{2}$ \and V.Yu. Avdeev$^{1}$ \and
O.V. Egorov$^{1}$}

\institute{Sternberg Astronomical Institute, Universitetskii pr.
13, Moscow, 119992 Russia \and Special Astrophysical Observatory,
RAS, Nizhnii Arkhyz, Karachai-Cherkessian Republic, 357147 Russia
}

\offprints{T.A.  Lozinsksaya, \email{lozinsk@sai.msu.ru}}

\date{}

\titlerunning{Structure and kinematics of the interstellar medium}

\authorrunning{Lozinskaya et al. }

\abstract{ The structure and kinematics of ionized gas in the
star-forming region in the BCD~galaxy VII~Zw~403 (UGC~6456) are
analyzed using observations with the \SCORPIO focal reducer on the
6-m Special Astrophysical Observatory telescope in three modes:
direct imaging (in the H$\alpha$, [OIII], and [SII] lines),
long-slit spectroscopy, and spectroscopy with a scanning
Fabry-Perot interferometer. In addition to the previously known
bright HII~regions and the faint giant ring that surrounds the
entire star-forming region, many new faint diffuse and arc
structures have been detected. Fine structure of the giant ring
has been revealed. We do not confirm the previously detected
expansion of the bright shells around young stellar associations
with a velocity of $50-70\km$. We have estimated their expansion
velocities to be no higher than 15--$20\km$; the corresponding
kinematic age, no younger than 3--4~Myr, agrees well with the age
of the compact OB~associations related to them. We associate the
faint extended filamentary and diffuse regions of ionized gas
identified  in nearly the entire central region of the galaxy and
the giant HII ring with the older (10~Myr) stellar population of
the most recent starburst. Weak high-velocity [OIII] and H$\alpha$
line wings (up to $300\km$ from the line center) have been
detected in the brightest HII~region. Such velocities have been
observed in the galaxy for the first time. The previously
published H$\alpha$ luminosity measurements for the galaxy are
refined.}

\maketitle

\section{Introduction}

Observations of giant shells and supershells around star clusters
of various ages in Irr and BCD~galaxies provide a unique
opportunity to study the interstellar medium in star-forming
regions at various epochs. Numerous observations of several nearby
galaxies suggest that  episodes of violent star formation in galaxies are
localized in space and in time and manifest themselves as giant
complexes where the bulk of the young stellar population, bright
HII regions, and shells and supershells of various scales are
concentrated. The ``local'' interaction of individual
OB~associations with the ambient interstellar gas shows up most
clearly in such complexes.

In this paper, we analyze in detail the structure and kinematics
of the shells and supershells that represent traces of several
sites of the most recent star formation episode in the BCD~galaxy
VII~Zw~403 (UGC~6456). This is a nearby, isolated, and slowly
rotating galaxy that has been studied extensively in the optical,
infrared, and X-ray. A study of the stellar population of
VII~Zw~403 with the Hubble Space Telescope (HST) allowed Lynds
et~al.~(1998) to perform multicolor photometry of individual stars
and to reveal several star formation episodes of various
intensities in the galaxy. According to these authors, the
strongest starburst occurred 800--600~Myr ago. Stars of this
generation dominate in the galaxy. Therefore, were it not for the
most recent (4--10~Myr ago), relatively weak starburst in the
central $\sim1$~kpc, the galaxy would be classified as a dwarf
elliptical~(dE). Loose and Thuan~(1985) classified VII~Zw~403 as
belonging to the most numerous class of BCD~galaxies, iE, which is
characterized by an irregular bright star-forming region near the
center of an extended elliptical halo of old stars. Near-infrared
photometry for the stellar population of VII~Zw~403 confirms that
the region of current star formation is surrounded by an extended
``old'' stellar halo (Schulte-Ladbeck et~al.~1999a).

The star-forming region is observed toward a giant cloud of
neutral hydrogen with the highest HI column density in the galaxy
and is surrounded by an extended HI halo $3\farcm6\times 2\farcm9$
or $4.7\times 3.8$~kpc in size (Thuan et~al.~2004). The youngest
and most massive stars form the distinct associations nos.~1, 2,
3, 4, 5, 6 (for uniformity, we will use the designations from
Lynds et~al. (1998)). The more evolved stars, including the most
luminous red supergiants, are distributed in a more extended
region and show no concentration to young clusters. All of the
ionized-gas emission is also concentrated in the central region of
the galaxy and reveals traces of several sites of the most recent
star formation episode: eight bright HII regions associated with
young stellar associations and weak diffuse emission surrounding
them (Thuan et~al. 1987). Based on HST observations, Lynds et~al.
(1998) identified a shell structure of several bright HII regions
70--150~pc in size. Silich et~al.~(2002) found traces of the faint
giant ($D\simeq 500$~pc) ring in the H$\alpha$~emission from the
diffuse gas. An extended region of diffuse X-ray emission inside
the giant ring was reported, but these data have not yet been
confirmed (Papaderos et~al. 1994; Lira et~al. 2000; Bomans~2001).
This question remains open: the observations by Ott et~al.~(2005a,
2005b) allowed only an upper limit for the diffuse X-ray
luminosity to be estimated, $L_{X} \leq 2.8\times 10^{38}~$
erg~s$^{-1}$.

The results of ionized-gas observations in the galaxy are
ambiguous. Thus, for example, the measurements of the total
H$\alpha$~luminosity of the galaxy, which is used as the main
criterion for the star formation rate, differ in the papers of
Lynds et~al.~(1998) and Silich et~al.~(2002) by more than a
factor of~20. Based on long-slit spectra taken with the 4-m KPNO
telescope, Lynds et~al.~(1998) found an expansion of bright
shells with a velocity $V_{exp} = 50-70\km$, but the observations
by Thuan et~al.~(1987) and Martin~(1998) with a higher spectral
resolution revealed no such high velocities in the galaxy. All of
these factors suggest that further observations of the ionized
gas in the star-forming region are required.

We observed the galaxy on the 6\mbox{-}m Special Astrophysical
Observatory (SAO) telescope with the \SCORPIO focal reducer in
three modes: direct imaging in medium-band filters, long-slit
spectroscopy, and spectroscopy with a scanning Fabry--Perot
interferometer. We will use our observations of the galaxy with
the integral field spectrograph MPFS to analyze in detail the gas
emission spectrum in our next paper (Lozinskaya et~al. 2007). In
this paper, we analyze the structure and kinematics of the ionized
gas in the star-forming region. In the Introduction, basic data on
the galaxy are gathered and the goals of our studies are
formulated. Next, the observations and data reduction techniques
are described, and the results obtained are presented and
discussed. In the final section, we summarize our main
conclusions.

In this paper, we assumed the distance to the galaxy to be
$d=4.5$~Mpc (Lynds et~al. 1998; Schulte-Ladbeck et~al. 1999b),
which corresponds to an angular scale of 22~pc~arcsec$^{-1}$.

\begin{table*}
%%% Table:1
\caption{Log of photometric observations}
\begin{tabular}{l|c|c|c|c|c}
\hline \multicolumn{1}{c|}{Range}&Date&$\lambda_{\textrm{c}}$, \AA&
\multicolumn{1}{c|}{FWHM, \AA}&$T_\textrm{exp}$,~s&Seeing \\
\hline
H$\alpha$        & Nov.~29/30, 2002& 6560  &  15    &\phantom{3}600          & $1\farcs2$ \\
H$\alpha$~continuum & Nov.~29/30, 2002& 6608  &  15    &\phantom{3}240          & $1\farcs2$ \\
H$\alpha$~$+$~[NII]  & May~25/26, 2003& 6552  &  75    &3000          & $1\farcs4$ \\
H$\alpha$~continuum& May~25/26, 2003& 6608  &  15    &1500          & $1\farcs5$ \\
{[OIII]}           & July~23/24, 2003& 4995& 220    &3600          & $1\farcs8$ \\
{[OIII]}~continuum& July~23/24, 2003& 5310& 170    &1800          & $1\farcs8$ \\
{[SII]}            & July~22/23, 2003& 6720&   15   &3000          & $1\farcs9$ \\
{[SII]}~continuum& July~23/24, 2003& 6604&  21    &1200          & $1\farcs6$ \\
\hline
\end{tabular}
\end{table*}

\section{Observations and data reduction}

\subsection{Interferometric Observations}

We performed our interferometric observations of VII~Zw~403 on
November~30, 2002, at the prime focus of the 6-m SAO telescope
(BTA) using the \SCORPIO focal reducer; the equivalent focal
ratio of the system was $F/2.9$. A description of \SCORPIO was
given by Afanasiev and Moiseev~(2005) and on the Internet
(\verb"http://www.sao.ru/hq/lsfvo"); the SCOPRIO capabilities in
interferometric observations were also described by
Moiseev~(2002).

The seeing during the observations varied within the range
$1\farcs5-2\farcs1$. We used a scanning Fabry-Perot
interferometer (FPI) operating in the 501st order at the
H$\alpha$ wavelength. The spacing between the neighboring orders
of interference, $\Delta\lambda=13$~\AA, corresponded to a region
free from order overlapping with a size of $\sim600\km$ on the
radial velocity scale. The width of the FPI instrumental profile
was $\sim0.8$~\AA, or $\sim35\km$. Premonochromatization was
performed using an interference filter with the FWHM
$\Delta\lambda =15$~\AA\ centered on the H$\alpha$ line. The
detector was a TK1024 $1024\times1024$-pixel CCD~array. The
observations were performed with $2\times2$-pixel binning to
reduce the readout time. In each spectral channel, we obtained
$512\times512$-pixel images at a scale of $0.55''$ per~pixel; the
total field of view was $4\farcm8$. We took a total of 36
interferograms for various FPI place spacings, so the size of the
spectral channel corresponded to $\delta \lambda=0.36$~\AA, or
$17\km$ near H$\alpha$. The exposure time was 300~s.

We reduced the raw data using software running in the IDL
environment (Moiseev~2002). After the primary reduction, the
observational data were represented as a $512\times512\times36$
data cube; here, a 36-channel spectrum corresponds to each pixel.
The final angular resolution (after smoothing during the data
reduction) was $\sim2\farcs2$. The smoothing was performed using
the ADHOC software package \footnote{The ADHOC software package
was developed by J.~Boulestex (Marseilles Observatory).}.

We constructed the radial velocity fields and the H$\alpha$
intensity and FWHM maps using single-component Gaussian fitting
of the emission line profile in the data cube. The FWHM map was
corrected for the instrumental profile using the ``standard''
relation (the square root of the difference between the squares
of the FWHMs). The formal accuracy of our radial velocity
measurements for symmetric line profiles was $\sim3-5\km$.

All of the radial velocities in this paper are heliocentric.

\subsection{Images in Emission Lines}

Emission-line and continuum images of VII~Zw~403 were obtained on
the 6-m SAO telescope with the \SCORPIO focal reducer. A log of
photometric observations is given in Table~1, which lists the
dates of observations, the central wavelengths
($\lambda_{\textrm{c}}$) and full widths at half maximum (FWHMs)
of the filters used, the total exposure times $T_\textrm{exp}$,
and the average seeing. This table lists the filters whose
passbands included the H$\alpha$,
(H$\alpha$+[NII]~$\lambda6548,6583$~\AA),
[OIII]$~\lambda4959,5007$~\AA, [SII]~$\lambda6717$~\AA\ lines and
the filters centered on the continuum near these lines.

Note that the observations in~2002 were performed with a narrow-band filter centered on
the H$\alpha$ line of the galaxy (this filter was also used for the FPI observations),
while the deeper 2003 image was obtained through a filter whose passband, in addition to
H$\alpha$, included the two neighboring [NII] lines.

In 2002 and 2003, the detectors were, respectively, a TK1024 CCD~array with a scale of
$0.28''$ per~pixel and an EEV42\mbox{--}40 CCD~array with a scale of $0.31''$ per~pixel
(in $2\times2$ readout mode). The data were reduced using a standard procedure; the
spectrophotometric standard stars AGK$+$81~266 and Grw$+$70~5824 were observed on the
same night and almost at the same zenith distance to calibrate the images on the energy
scale.

When constructing the maps of ``pure'' line emission, we
subtracted the images in the corresponding continuum from the
images in the filters containing the emission lines. In the
subtraction, the normalization factor was chosen in such a way
that the images of field stars were mutually subtracted in the
resulting frame. Since VII~Zw~403 is a galaxy with a blue
continuum (at any rate in the star-forming region) and since the
field stars are mostly red, this (generally accepted) procedure
could lead to an underestimation of the contribution from the
real continuum for the [OIII] and H$\alpha$ lines (here, we took
a redder filter for the continuum) and, conversely, to an
overestimation of the contribution from the continuum for the
[SII] images. This effect is very difficult to properly take into
account (we do not know the slope of the spectral continuum in
different regions of the galaxy and in field stars), which should
be remembered when considering the line-ratio maps. For an
additional check, we used preliminary spectrophotometric MPFS
data for the central region of the galaxy, since the continuum in
these observations was subtracted without problems. Unfortunately,
 our MPFS
data are available only for the central region of the galaxy.

\subsection{Long-Slit Spectroscopy}

Spectral observations of VII~Zw403 were performed with the same
\SCORPIO instrument operating in the mode of a slit spectrograph
with a slit about $6'$ in length and $1''$ in width. The scale
along the slit was $0.36''$ per~pixel. We used volume phase
holographic gratings (VPHGs) that provided a spectral resolution
of $130-140\km$ near the [OIII] and H$\alpha$ lines. Since the
main objective of the observations with a slit spectrograph was
to verify the splitting of the H$\beta$ emission line in bright
shells detected by Lynds et~al.~(1998), we took two spectra with
a similar slit localization: scan~1 through regions N2, N3, N4
and scan~2 through the HII regions N1 and N5 (in the designations
of Lynds et~al. 1998). The position angle of the spectrograph
slit was $88^\circ$ in both cases. In addition to the exposure
time $T_{\textrm{exp}}$ and the average seeing, the log of
observations (Table~2) also contains the spectral range
$\Delta\lambda$.

\begin{table}
\caption{Log of spectral observations}
\begin{tabular}{c|c|c|c|c}
\hline
Scan&Date&$\Delta\lambda$, \AA&$T_{\textrm{exp}}$, s&\parbox[c][1cm]{2.1cm}{Seeing}\\
\hline
1 & 21/22.12.03& 6270--7300 & 1200 & $2\farcs2$ \\
1 & 05/06.12.04& 4800--5600 & 1200 & $2.0$ \\
2 & 21/22.12.03& 6270--7300 & 1800 & $2.2$ \\
2 & 05/06.12.04& 4800--5600 &\phantom{1}800 & $2.0$ \\
\hline
\end{tabular}
\end{table}
We took the spectra not only in the ``green'' spectral range containing the H$\beta$ and
[OIII] $\lambda4959$, 5007~\AA\ lines, as in Lynds et~al.~(1998), but also in the
``red'' spectral range including the H$\alpha$, [NII] $\lambda6548$, 6583~\AA, and [SII]
$\lambda6717, 6731$~\AA\ lines. The data were reduced in a standard way; the
spectrophotometric standard AGK$+$81~266 observed immediately after the object at a
close zenith distance was used for energy calibration.

The radial velocities and FWHMs of the emission lines were
determined by single-component Gaussian fitting of their
profiles; the observed line FWHMs were corrected for the
instrumental profile, as in the case of the FPI data. The
accuracy of the absolute velocity measurements was checked using
night-sky lines and was $\sim5\km$.

\begin{figure*}
\includegraphics[width=17 cm]{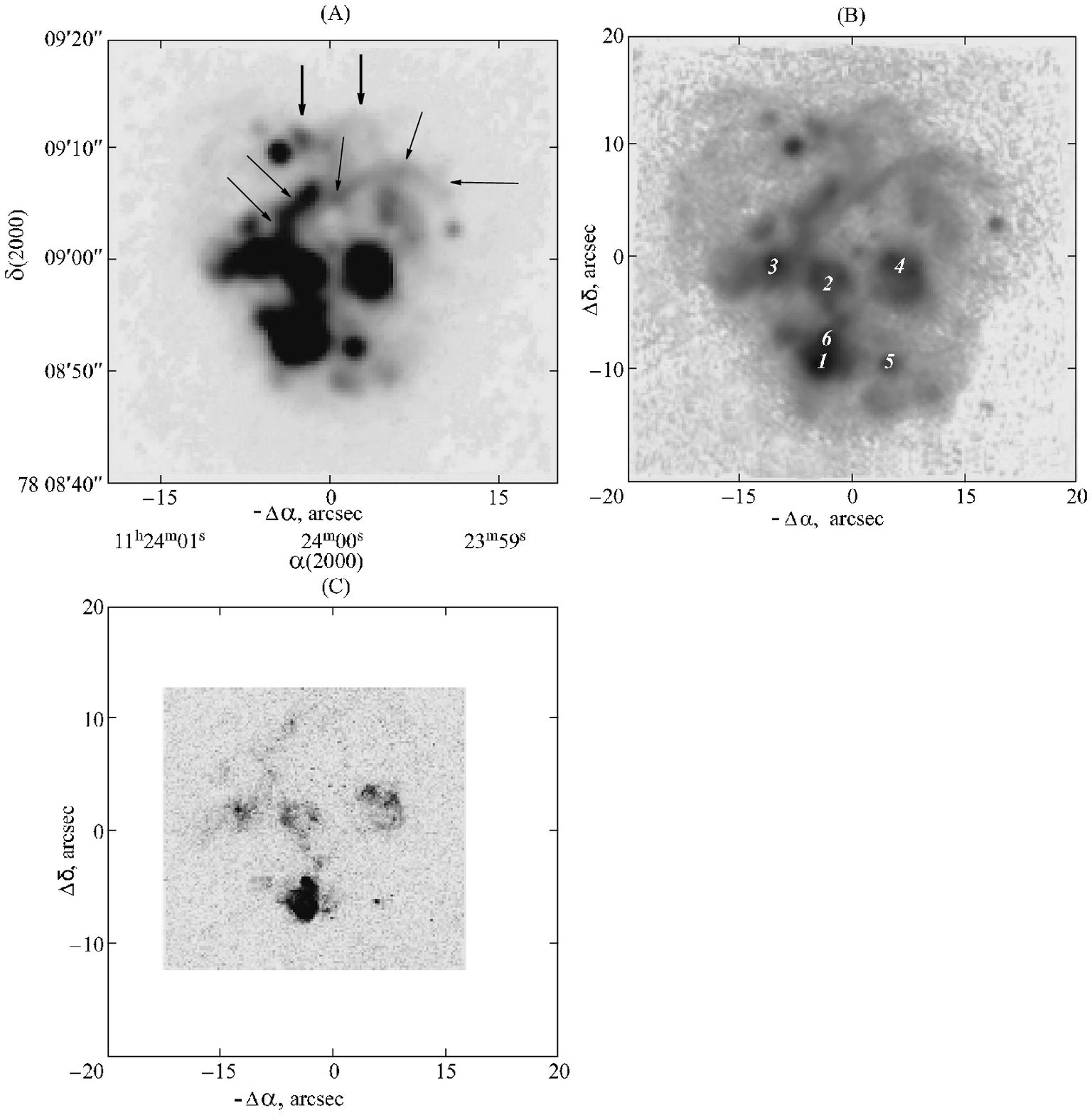}
\caption{H$\alpha$~images of the star-forming region: the images
with the \SCORPIO focal reducer of the 6-m SAO telescope on (a)
linear and (b) logarithmic scales and (c) the WFPC2 image from
the HST archive in the F656N filter. The thin arrows indicate the
individual arcs which the giant ring identified by Silich et~al.
(2002) consists of; the thick arrows indicate the outer arc
outside this ring. The numbers mark the HII regions identified by
Lynds et~al.~(1998).}
\end{figure*}

\begin{figure*}
\centerline{\includegraphics[width =8.5 cm]{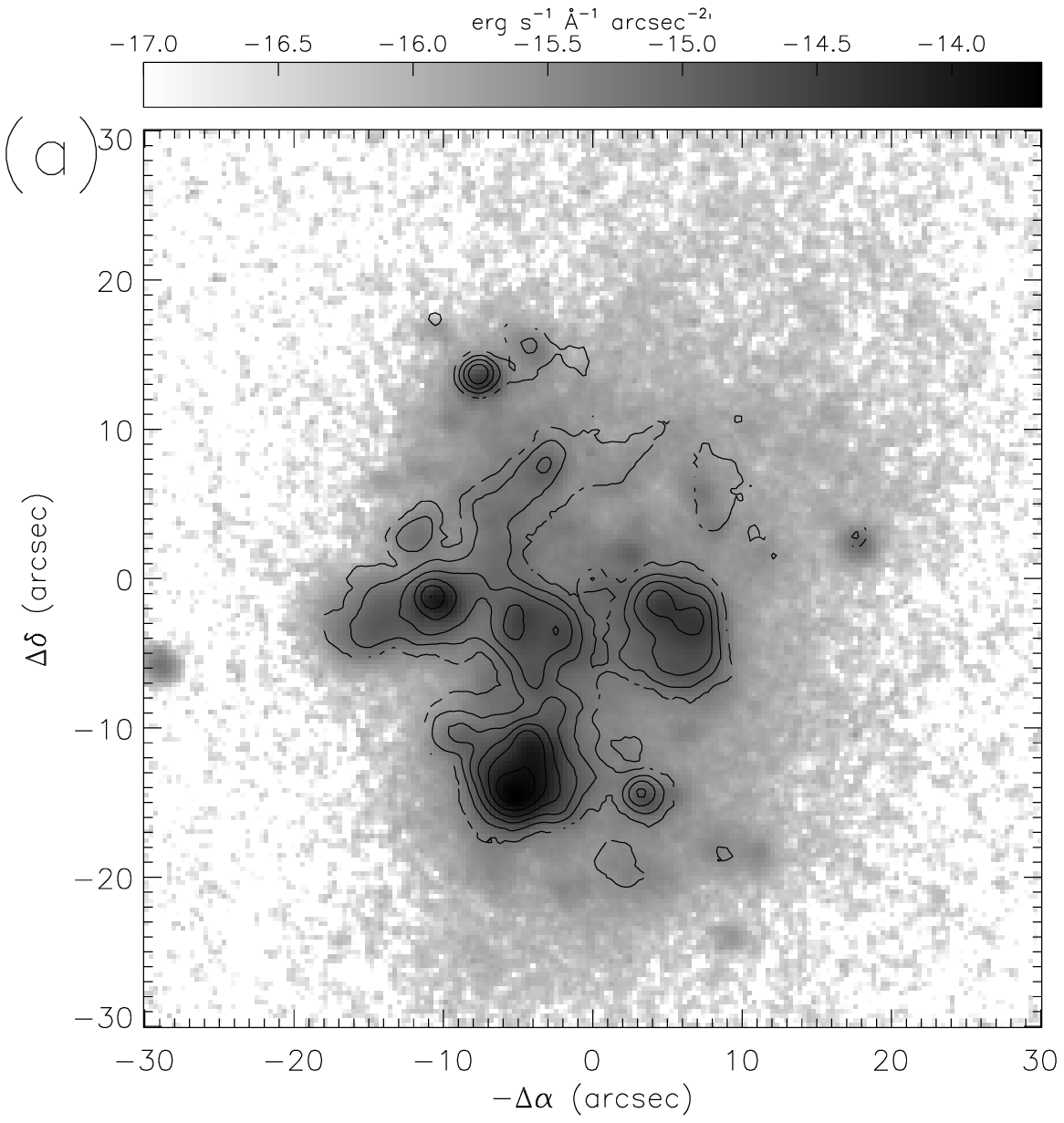}
\includegraphics[width =8.5 cm]{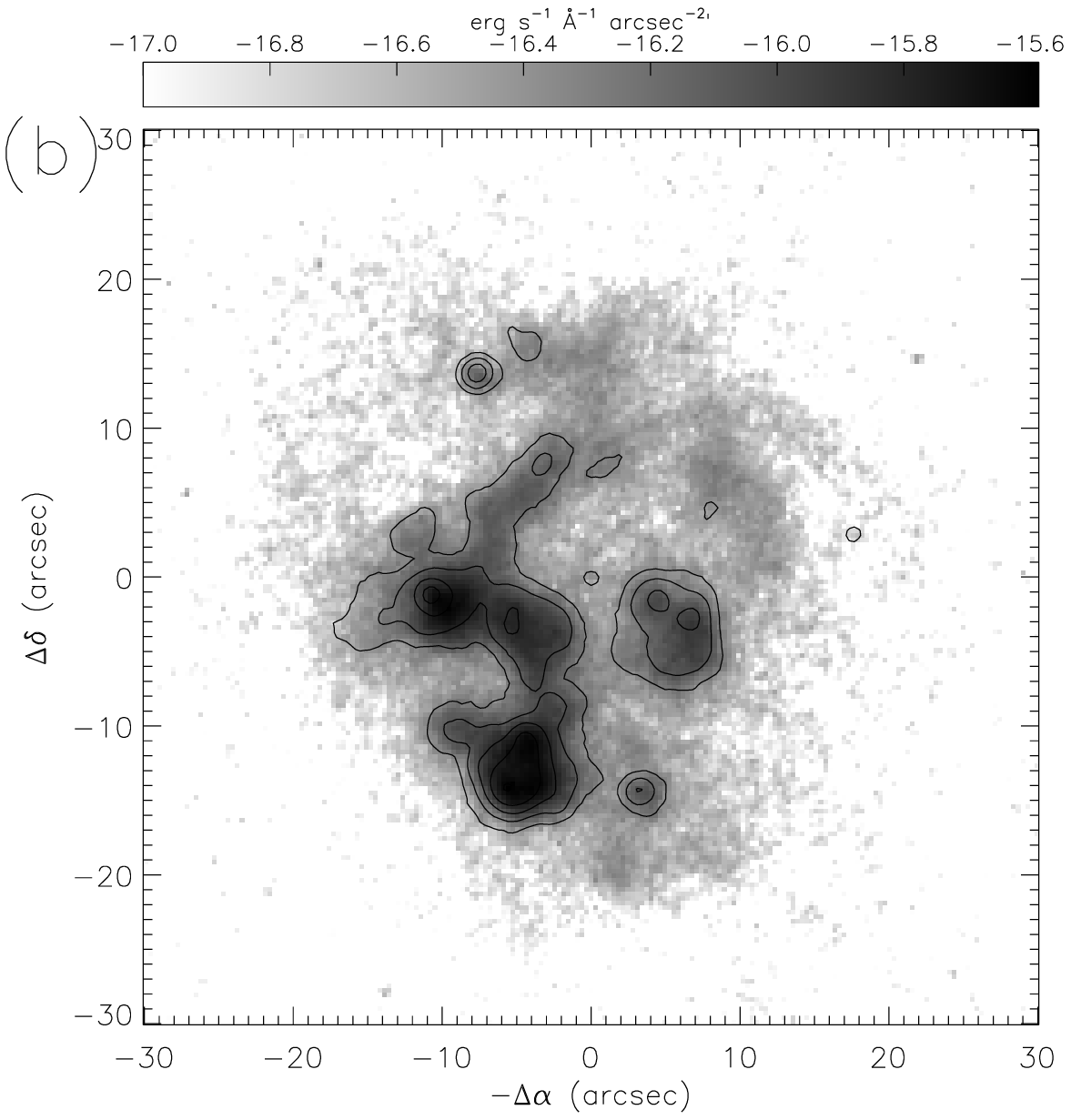}}
\centerline{\includegraphics[width =8.5 cm]{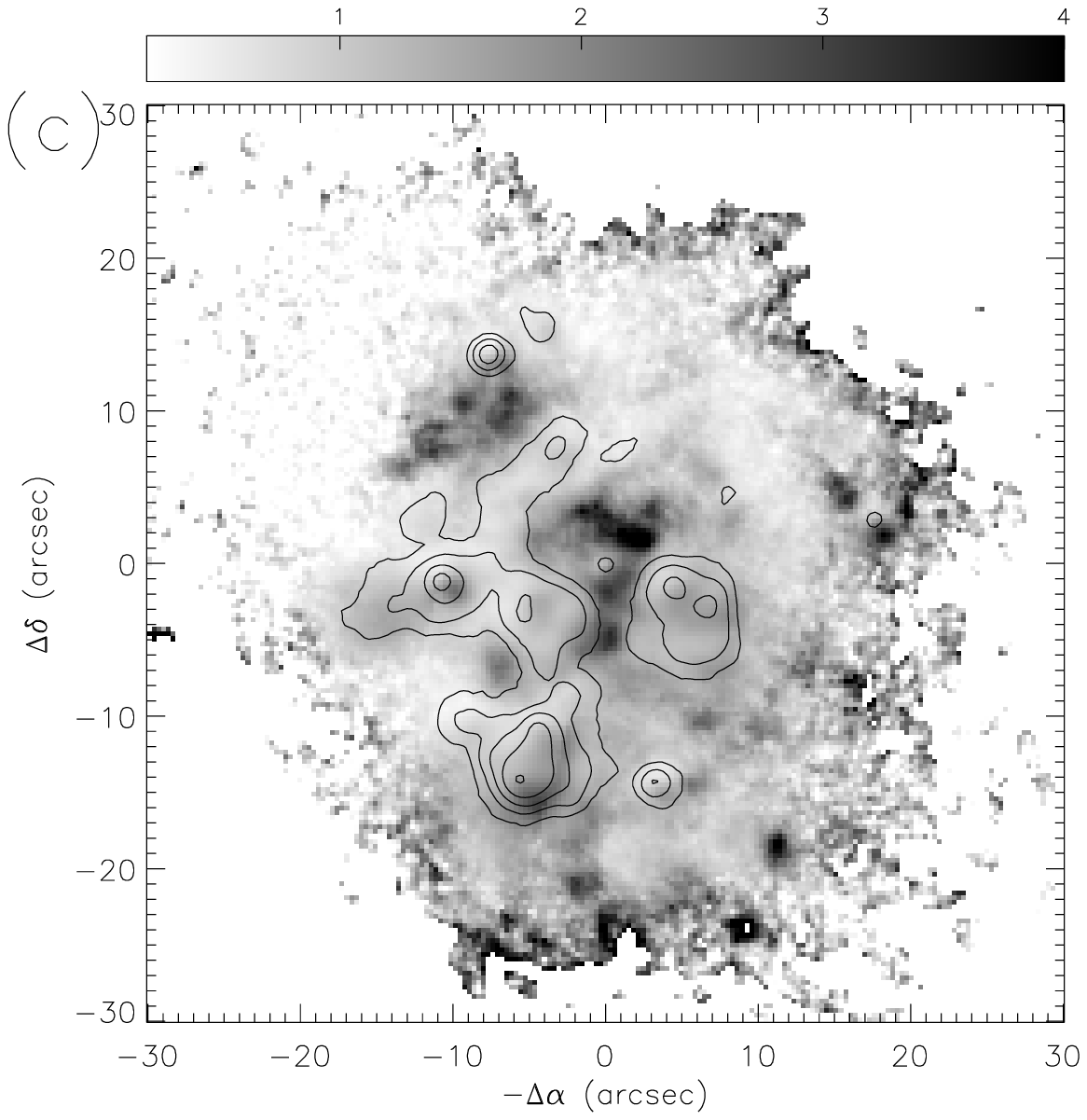}
\includegraphics[width =8.5 cm]{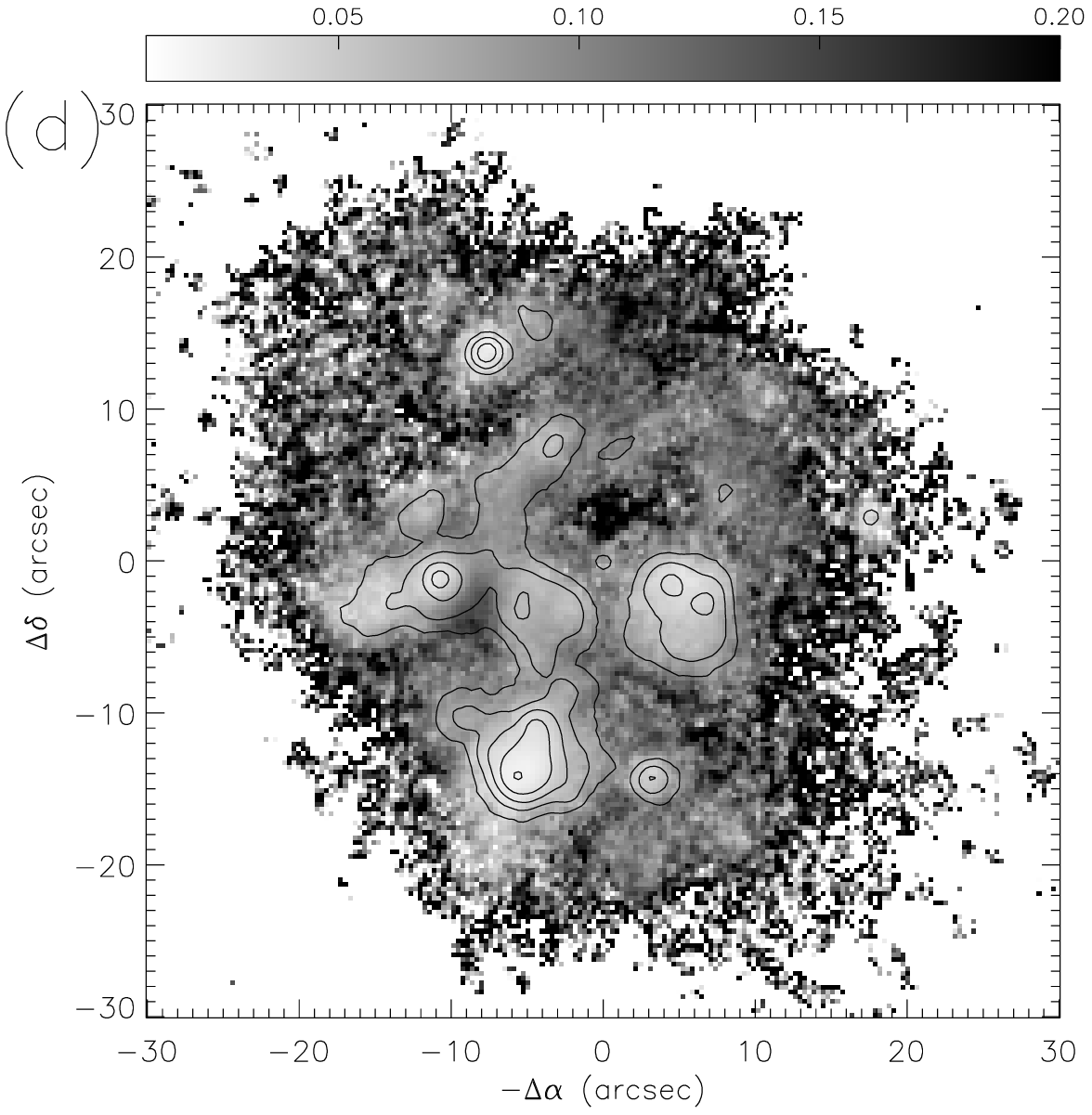}}
\caption{(a) [OIII] and (b) [SII] images of the star-forming
region on a logarithmic scale with the H$\alpha+$[NII] line
contours superimposed and the  (c) [OIII]/(H$\alpha+$[NII]) and
(d) [SII]/(H$\alpha+$[NII]) ratio maps with the same contours
superimposed. \hfill}
\end{figure*}

\section{Results}
\subsection{General Structure of the Star-Forming Region in the Galaxy}

Direct medium-band images of the central region of the galaxy in
the (H$\alpha$ + [NII]), [SII] 6717\AA, and [OIII] 5007\AA\ lines
were obtained with the \SCORPIO focal reducer of the 6-m SAO
telescope.

\begin{table*}
\caption{Estimates of the H$\alpha$ flux and luminosity for VII~Zw~403}
\begin{tabular}{l|c|l}
\hline
\multicolumn{1}{c|}{Reference}& Flux, erg~s$^{-1}$~cm$^{-2}$ & \multicolumn{1}{c}{$L$, erg~s$^{-1}$, $D=4.5$~Mpc} \\
\hline
Lynds et~al.~(1998) & --  & $1.80\times 10^{39}$ ($\Ha$)    \\
Martin~(1998)       &$6.5\times 10^{-13}$  &  $1.58\times10^{39}$~(\Ha+[NII])  \\
Silich et~al.~(2002)& $1.95\times 10^{-11}$&  $4.70\times10^{40}$~(\Ha)    \\
BTA    (15A filter) & $6.14\times 10^{-13}$&  $1.49\times10^{39}$~(\Ha)      \\
BTA    (75A filter) & $7.96\times 10^{-13}$&  $1.93\times10^{39}$~(\Ha+[NII]) \\
\hline
\end{tabular}
\end{table*}

Figure~1 shows the BTA H$\alpha$ images of the star-forming region on linear and
logarithmic scales and its image in a filter containing the H$\alpha$ line from the HST
archive. The numbers in Fig.~1b mark the HII regions identified by Lynds et~al.~(1998).

As follows from this figure, all of the HII shells N1--N6
identified with HST are seen as the brightest emission regions in
our images (with a lower angular resolution that does not allow
their shell structure to be revealed). In addition to these
previously known bright HII regions, our images reveal a large
number of new, fainter emission features. These include
spherically symmetric  regions ($2-5''$ in  size), extended arc
and linear structures, and a large region of weak diffuse
emission.

Our observations show that the giant shell structure 250~pc in radius torn in the
southwest (below referred to as the giant ring) that was identified by Silich
et~al.~(2002) consists of several individual arc structures with different brightnesses
and different radii of curvature (see Fig.~1a). Another outer arc and three HII regions
about $2''$ in size can be identified in the north outside it.

Figure~2 allows the large-scale structures of the central star-forming region in
different lines to be compared. It shows the [OIII] and [SII] images on a logarithmic
scale with the H$\alpha+$[NII] line contours superimposed and
the [OIII]/(H$\alpha+$[NII]) and
[SII]/(H$\alpha+$[NII]) line ratio maps with the same contours superimposed.

As follows from Figs.~1 and~2, in general, the large-scale structures of the star-forming
region in the (H$\alpha +$~[NII]), [SII], and [OIII] lines are almost identical.
Nevertheless, two extended faint regions of diffuse emission in which the
[OIII]/(H$\alpha+$[NII]) intensity ratio is enhanced can be distinguished. One of these regions
lies inside the giant ring between the bright regions N2 and N4 to the north of them,
while the second region lies outside the giant ring between it and the outer arc in the
north (see Fig.~2c).

The [SII]/(H$\alpha+$[NII]) intensity ratio is also enhanced in the first region (see Fig.~2d).

The total size of the region of diffuse ionized gas in the galaxy
with all of the bright HII regions observed on its background is
$35-40''$ or $\sim900$~pc at a brightness level of $5\times
10^{-17}$~erg~s$^{-1}$~cm$^{-2}$~arcsec$^{-2}$.

We estimated the total H$\alpha$ flux from the central region of VII~Zw~403 by
integration within a circular aperture $40''$ in diameter. The fluxes and the
corresponding luminosity of the galaxy are given in Table~3. The correction for
extinction was made for the mean value of $E(B-V)=0.04$ (and $R_{v} = A_{v}/E(B-V) =
3.1$) found by Lynds et~al.~(1998) from all stars of the galaxy. We corrected the data
from Martin~(1998) in the same way; the luminosity is given for a distance of 4.5~kpc.
Silich et~al.~(2002) applied a correction only for the background extinction
$E(B-V)=0.02$, as inferred by Burstein and Heiles~(1984); the correction for internal
extinction in VII~Zw~403 increases further the flux and luminosity found by these
authors.

As follows from Table~3, our measured H$\alpha$ luminosity of the galaxy is considerably
lower than the value obtained by Silich et~al.~(2002) and agrees well with the previous
estimates by Lynds et~al.~(1998) and Martin~(1998).

\begin{figure*}
\centerline{\includegraphics[width =8.5 cm]{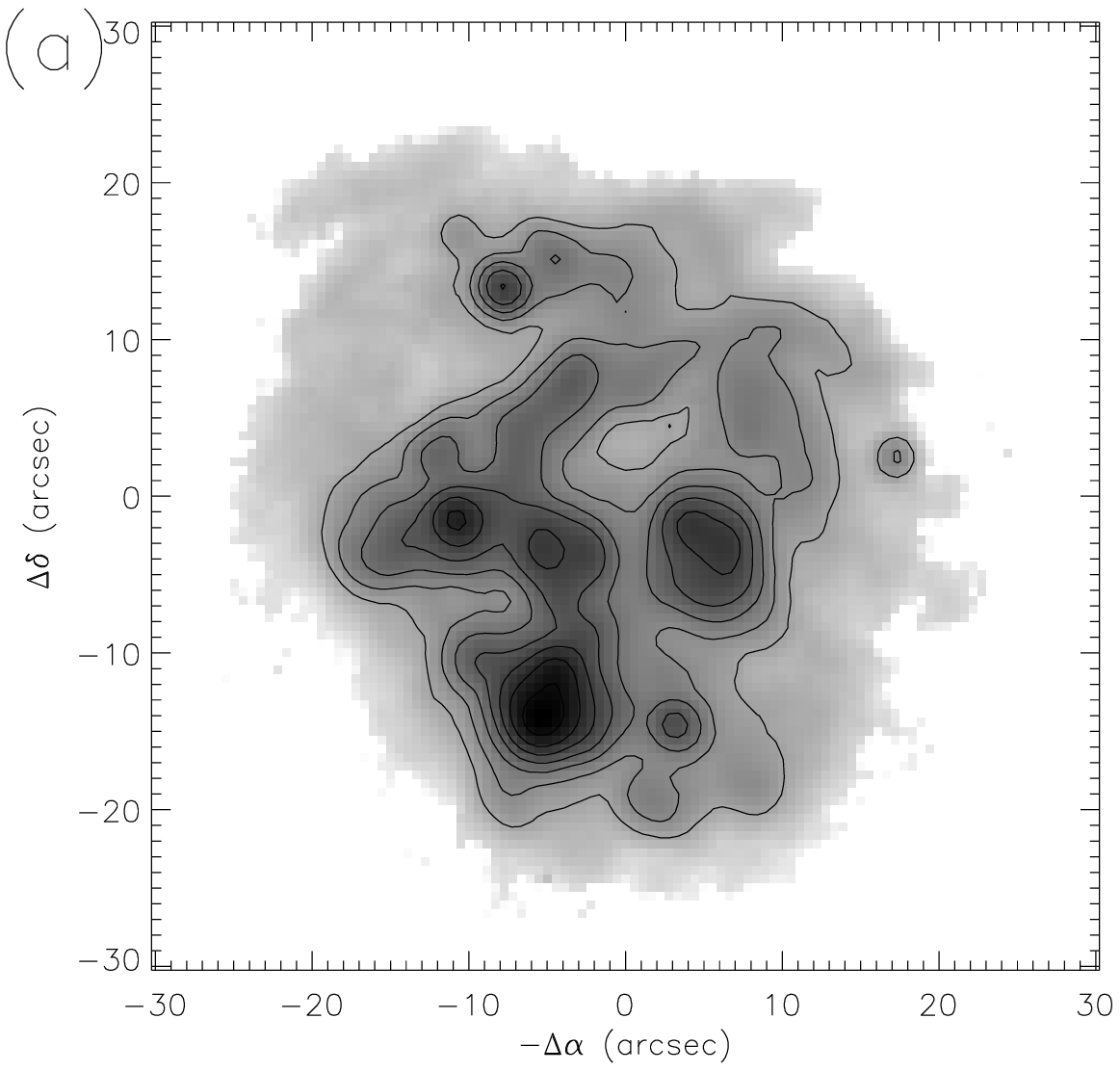}
\includegraphics[width =8.5 cm]{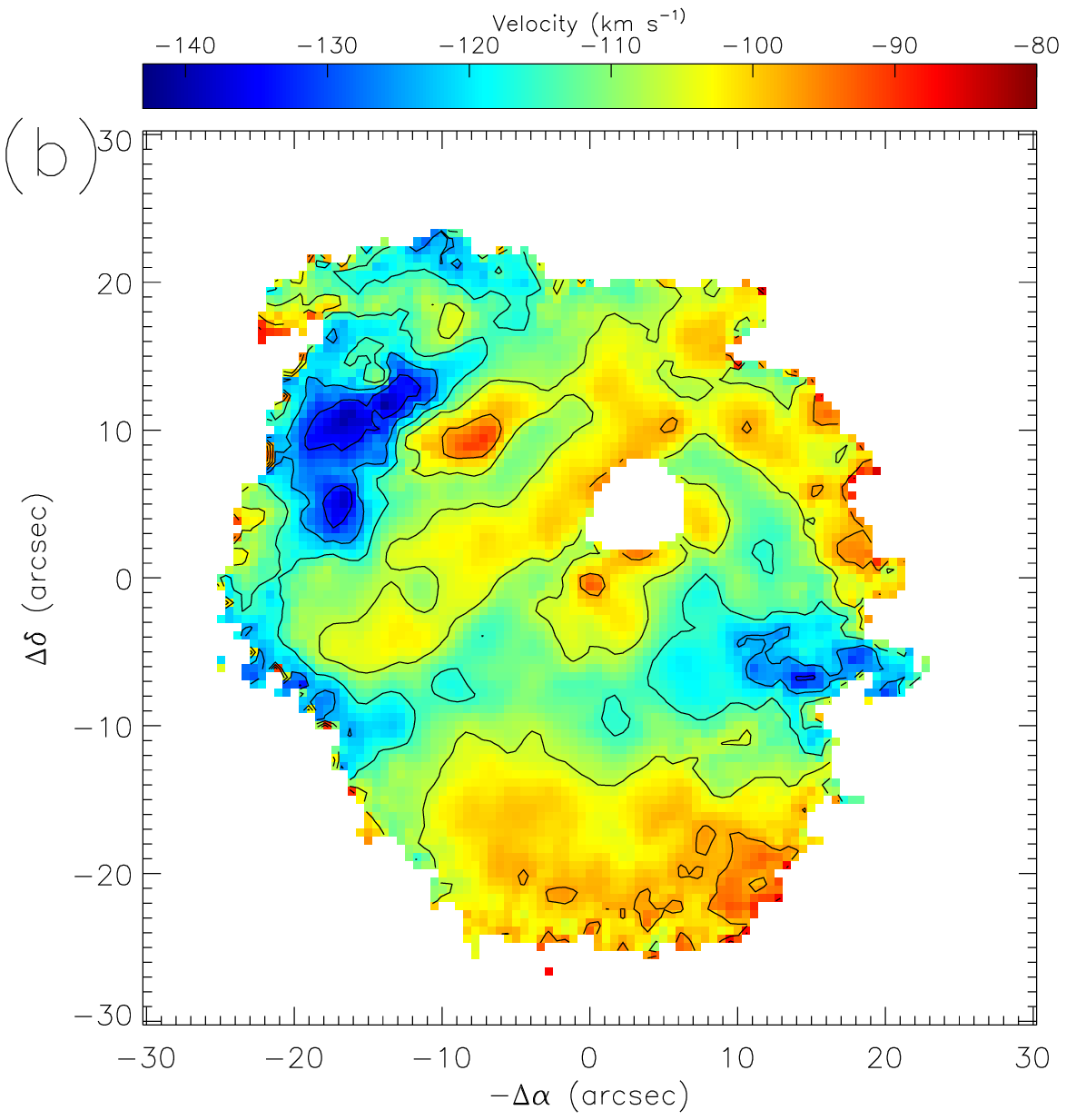}}
\centerline{\includegraphics[width =8.5 cm]{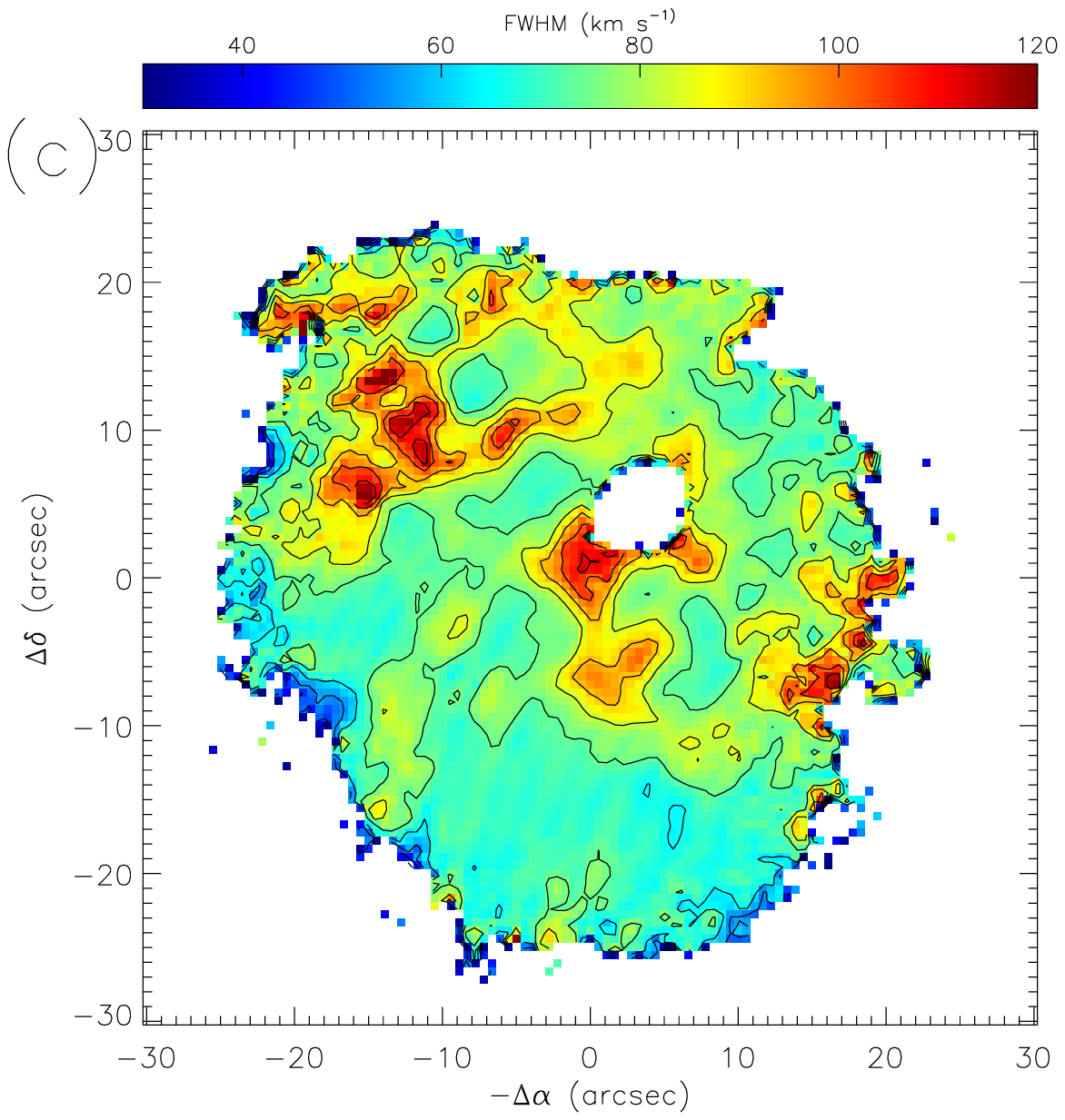}}
\caption{H$\alpha$ intensity (a), radial velocity (b), and FWHM
(c) distributions constructed from the FPI observations. The
``hole'' in the central region results from a parasitic ghost
inside the instrument. \hfill}
\end{figure*}

\begin{figure*}
\includegraphics[width=17 cm]{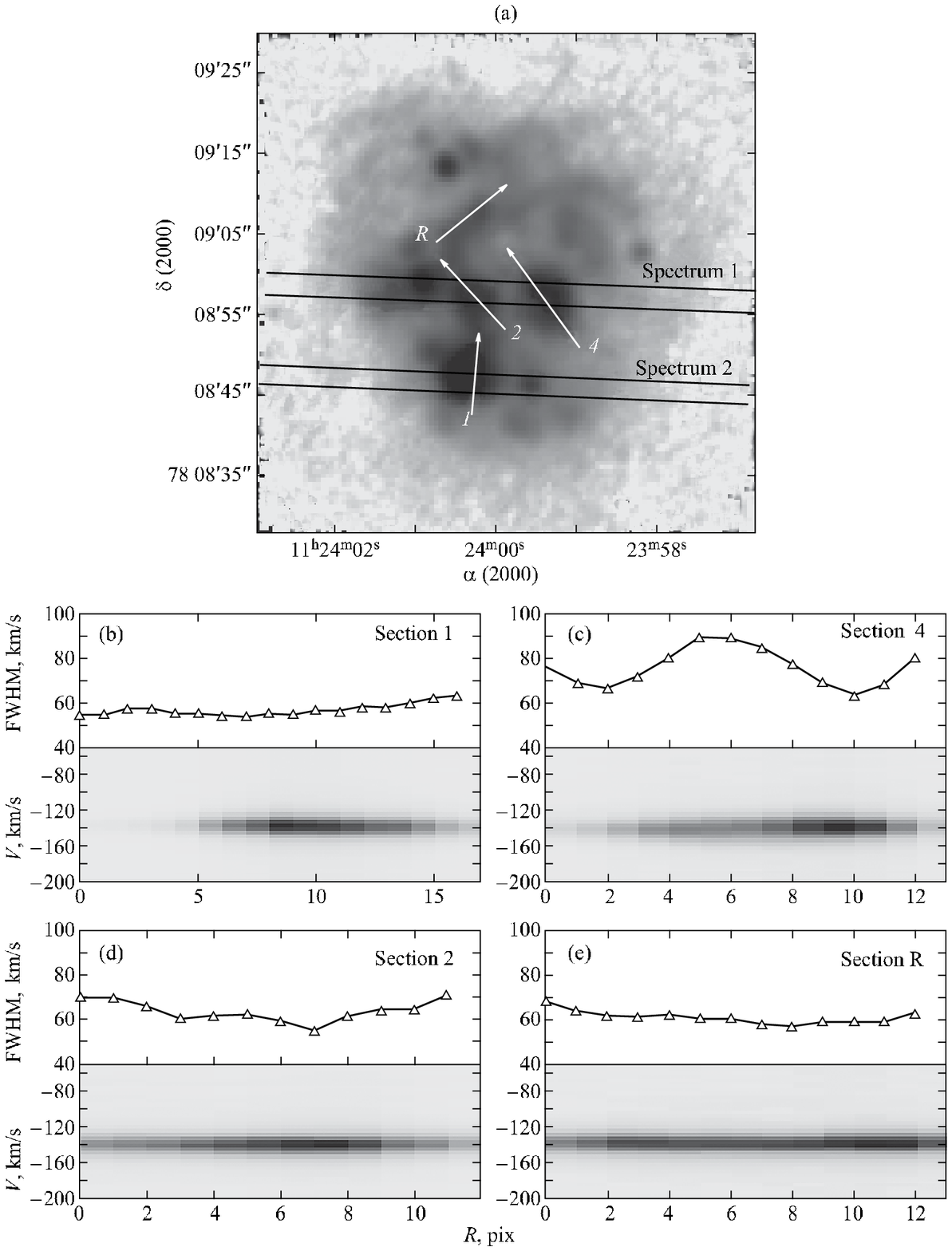}
\caption{(a) Orientation of the scans chosen as an example
together with the spectrograph slit localization for the two
spectra obtained. (b)--(e) Examples of the position--velocity
($P/V$) and position--FWHM ($P/\Delta V$) diagrams constructed
from the FPI observations.}
\end{figure*}

\begin{figure*}
\includegraphics[width=17 cm]{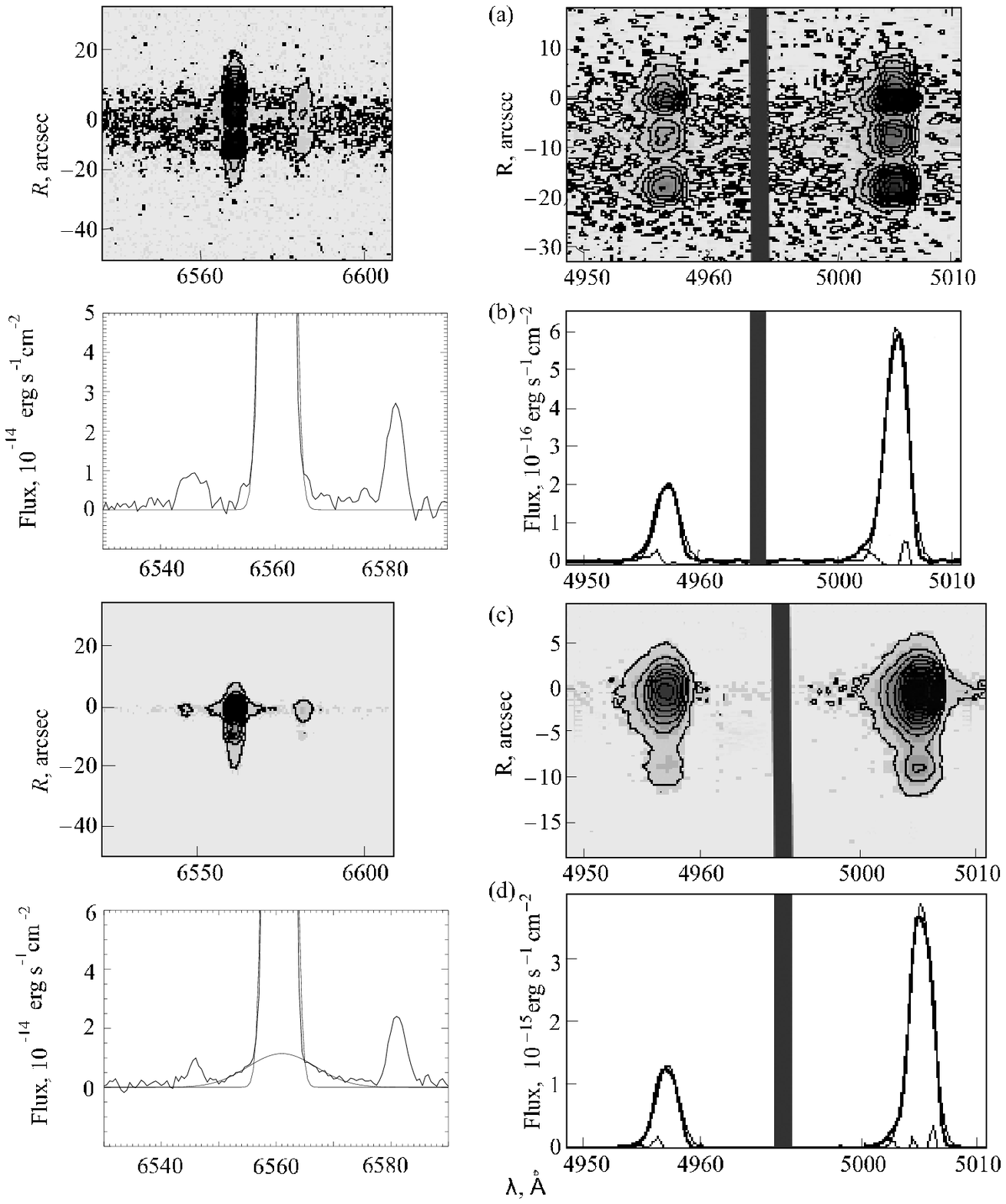}
\caption{Spectra around H$\alpha$ (on the left) and [OIII] (on
the right) lines at various intensity levels and their Gaussian
fits for two spectra: (a, b) spectrum~1 and (c, d) spectrum~2.
The localization of the spectrograph's slits  is shown in Fig.~4a.
\hfill}
\end{figure*}

\begin{figure*}
\centerline{\includegraphics[width =8.5 cm]{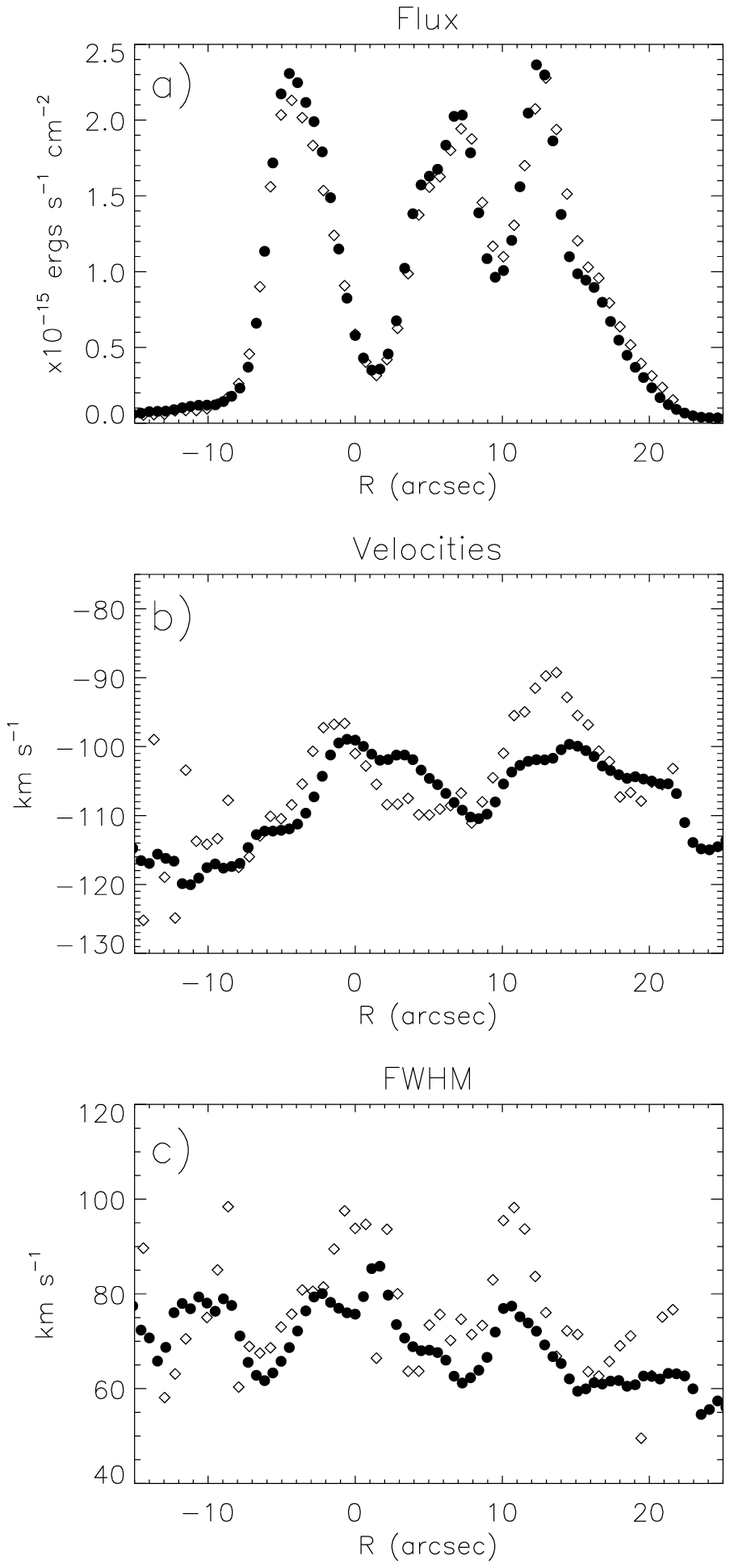}
\includegraphics[width =8.5 cm]{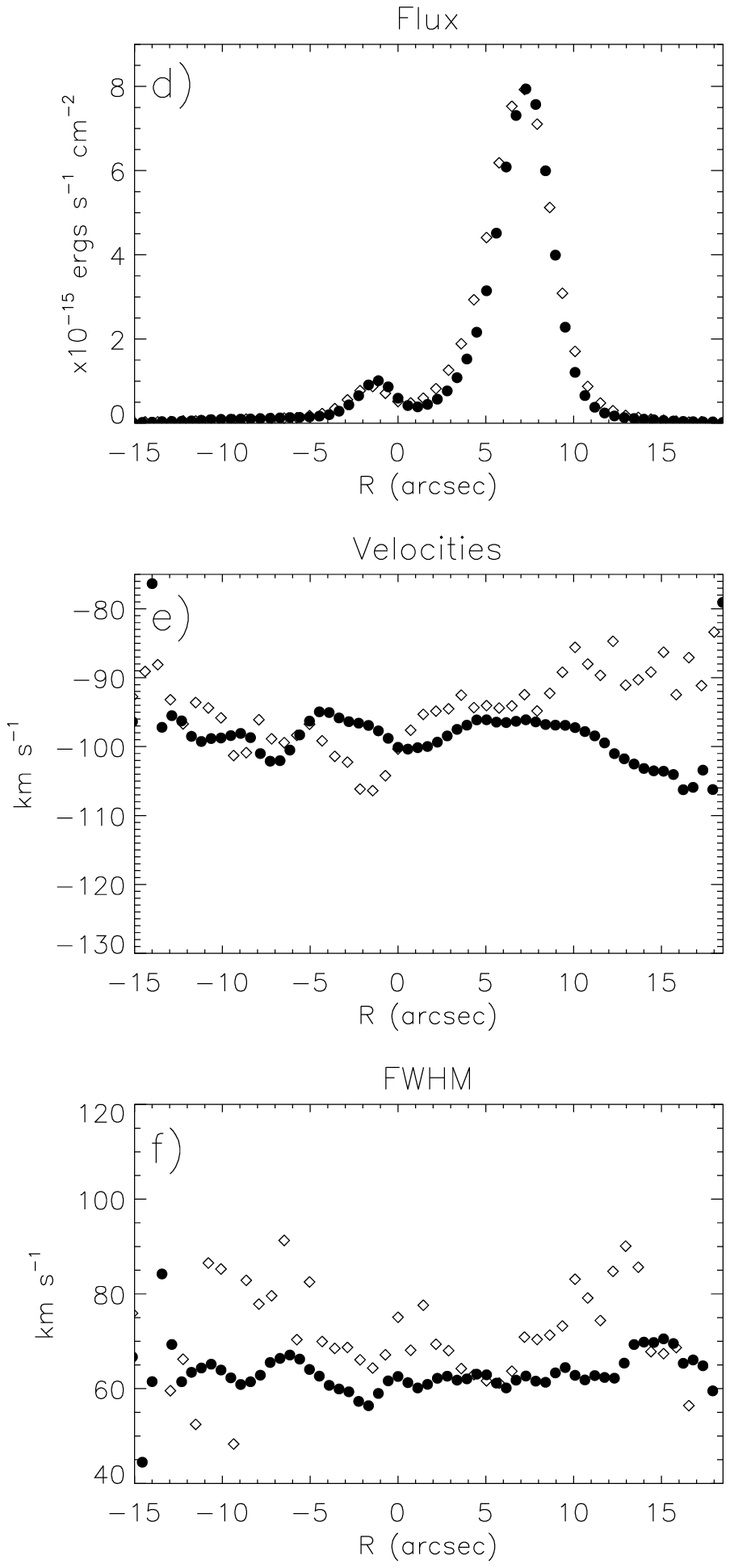}}
\caption{Comparison of the results of our spectral (open diamonds)
and interferometric (black circles)
observations: H$\alpha$ intensity, radial velocity, and FWHM variations along the two
scans corresponding to the spectrograph slit position shown in Fig.~4a: (a, b, c)
spectrum~1 and (d, e, f) spectrum~2.}
\end{figure*}

\subsection{Gas Kinematics in the Star-Forming Region}

\subsubsection{FPI observations}

Based on the observations with FPI, we constructed the H$\alpha$ radial velocity and
FWHM distributions in the central region of the galaxy (see Fig.~3).

The derived velocity of the line peak in the star-forming region
ranges from $-120$ to $-90\km$ (reaching $-130$...$-140\km$ in the
region of weak emission at the edge of the field). These
velocities agree well with the neutral-gas velocity from $-120$
to $-80\km$ in the extended surrounding region of the galaxy and
from $-110$ to $-90\km$ in the densest HI~cloud in the galaxy
with which the star-forming region coincides (Thuan et~al. 2004).

The H$\alpha$ FWHM corrected for the instrumental profile lies
within the range $60-70$ to $100-120\km$.

To find the expansion of the bright spherically symmetric
HII~regions, we constructed the so-called position--velocity
($P/V$) and position--FWHM ($P/\Delta V$) diagrams. The
orientation of the scans was chosen in such a way that they
crossed in two or three different
directions each of the HII regions, including those that reveal a
well-defined shell structure in the HST images (the HII regions
N1, N3, and N4) and whose expansion with velocities of $50-70\km$
was reported by Lynds et~al.~(1998).

As an example, Fig.~4 shows four of more than thirty our scans.

The most extended shell~N4 $6.5-8''$ (140--180~pc) in size showed
an increase in the line FWHM toward the center to $\Delta
V=90\km$ compared to $\Delta V=65-70\km$ in the bright northern
and southern peripheral parts. This may be the result of its
expansion with a velocity of $15-20\km$ (see scan~4 in Fig.~4).

According to our FPI observations, none of the bright HII~regions, except~N4, revealed
the velocity-ellipse structure characteristic of an expanding thin shell in the
$P/V$~diagrams or line broadening at the center in the $P/\Delta V$~diagrams.

In the region of the richest OB~association and the brightest HII
shell~N1, the HST image shows a chain of associations and
HII~regions associated with them: supershell~N1 (around
association no.~1), the HII~region N6 (around association no.~3),
and another region associated with association no.~2 between
them. (To avoid confusion , we retained all of the designations
from Lynds et~al. (1998)). Scan~1 in Fig.~4 passes along this
chain; our observations revealed no changes in the line velocity
or FWHM in this chain of HII~regions.

The line FWHM in the bright shells showing no evidence of
expansion is $60\km$, which exceeds the contribution from the
thermal and turbulent velocities in HII~regions. Of course, at an
angular resolution of $2\farcs2$, we average all of the radial
velocity variations over a region comparable in radius to the
shell. However, in any case, the hypothetical expansion velocity
of these shells that could explain the observed line FWHM does
not exceed $15-20\km$.

Based on the FPI observations, we determined the mean line
velocities and FWHMs in all of the faint H$\alpha$ emission
regions that were first identified in our images. No appreciable
line velocity and FWHM variations indicative of expansion with a
velocity higher than $20\km$ were found in any of them either.

To summarize the results of our kinematic studies with FPI, we
conclude that no sharp variations of the H$\alpha$ velocity and
FWHM are observed in the entire star-forming region on scales
larger than 45--50~pc, which corresponds to the actual angular
resolution of $2\farcs2$ for our FPI observations. The observed
velocity variations lie within the range $\pm(10-20)\km$; the line
FWHM ranges from~60 to $100-120\km$.

The accuracy of our measurements is high enough for the fast shell expansion reported by
Lynds et~al. (1998) to be detected. Thus, for example, the constructed FPI maps clearly
show a relationship between the HII~regions and the variations in the radial velocities
and velocity dispersion of the gaseous clouds (line FWHM).

\subsubsection{Long-slit spectroscopy}

To elucidate the cause of the discrepancies in  the expansion
velocities of the bright shells estimated from the spectral
observations by Lynds et~al.~(1998) and from our FPI
observations, we performed long-slit spectroscopy in the red
(H$\alpha$, [NII], [SII]) and green (H$\beta$, [OIII]) spectral
ranges. The spectral observations were performed under the
conditions closest to those in the observations by Lynds
et~al.~(1998). The actual spectral resolution (estimated from sky
lines) was FWHM=$120\km$, which is comparable to that in the
above work. The angular resolution was also similar, $2''$; it
was limited by seeing in our work and equal to the spectrograph
slit width in Lynds et~al~(1998).

In both ranges, we made two scans through the same regions of the shells whose spectra
are presented in Lynds et~al.~(1998). The spectrograph slit localization is shown in
Fig.~4a.

Figure~5 shows the H$\alpha$ and [OIII] lines in each of the two scans at
different intensity levels and Gaussian fits of their  profiles.

No distinct profile splitting that would be indicative of the
expansion of the ionized supershells with a velocity of
$50-70\km$ was found in any of the lines in the spectra. A
single-component Gaussian fits well the profiles of the H$\beta$,
H$\alpha$, and forbidden lines in the bright supershells N2, N3,
N4, and N5; the residuals are small.

Very weak high-velocity [OIII] and H$\alpha$ line wings are
observed in the brightest shell~N1 around the richest and
youngest association no.~1. As we see from Figs.~5c and~5d, the
[OIII] line here is asymmetric and reveals a weak blue component
with velocities of $-200$..$-300\km$ from the line center at
$\sim5\%$ of the maximum. The H$\alpha$ line exhibits a weak
broad pedestal that can be fitted by a Gaussian with
FWHM=$630\km$; the total flux in the broad component is $\sim4\%$
of the total flux in the line.

Such high velocities have been detected in the galaxy for the
first time and, in all probability, are indicative of gas
acceleration at the shock front. Of course, given the very low
intensity of the high-velocity features, these results must be
confirmed by independent observations. We performed several tests
to check whether these spectral features are instrumental effects
(a complex structure of the instrumental contour of the
spectrograph, focusing peculiarities, etc.). In particular, we
checked the shape of the instrumental profile using both intense
sky lines and a calibration comparison spectrum and made sure
that the instrumental profile contained no broad wings, at any
rate above $0.5-1\%$ of the maximum. Note also that the relative
intensity of the broad pedestal exceeds the intensity of the
scattered light in the \SCORPIO spectrograph with VPHG~gratings
(according to our estimates, less than $1\%$).

Figure~6 compares the results of our spectral and interferometric
observations: the H$\alpha$ intensity, radial velocity, and FWHM
variations along scans~1 and~2 whose localization is shown in
Fig.~4a (spectra~1 and~2).

First of all, this figure is an independent test of the reduction
of the data from the two observing programs for correctness: as
we see, the agreement between both the radial velocities and the
FWHMs of the lines derived from our interferometric and spectral
data is satisfactory. As we noted above, no line broadening at
the center is seen in the bright HII~regions, as was expected for
shells expanding with velocities of $50-70\km$. In contrast, a
clear line broadening  (from FWHM$\simeq60-70\km$ to FWHM $\simeq
80-120\km$) is observed in the faint region of diffuse ionized
gas between the bright HII~regions N4 and N2 as well as between
regions N3 and N2. This effect is also present in the second scan
crossing regions N1 and N5, but it is less prominent.

Our observations in the [OIII] line also revealed a similar line broadening in the
region of weak diffuse emission compared to the bright HII~regions. Thuan et~al.~(1987)
also pointed out that the bright HII~regions are characterized by the median radial
velocities, while the minimum and maximum H$\alpha$ velocities are observed in the weak
diffuse emission between them.

The large-scale neutral-gas kinematics in the star-forming region
reveals a distinct gradient in the radial velocity of the 21-cm
line peak (see Fig.~20 in Thuan et~al. 2004). In the direction
from northeast to southwest, the velocity smoothly increases from
$-110$ to $-95--85\km$ in both HI~clouds elongated in this
direction. The nature of this gradient is unclear; the above
authors suggest that the gas may rotate in the central region
along the major axis of the galaxy. We attempted to find the same
gradient in the ionized-gas velocity distribution in the region.
To this end, we constructed from our FPI H$\alpha$ observations
 $P/V$ diagrams in the directions
identical to those of the $P/V$ diagrams for~HI in Thuan et~al.
(2004). We found no similar
smooth gradient in the H$\alpha$ radial velocity.

\section{Discussion}

Our (H$\alpha$~$+$~[NII]), [SII], and [OIII] images for the
central region of the galaxy suggest that the large-scale
structures of the star-forming region in these lines are
identical. All of the previously identified HII~regions are seen
as the brightest emission features in our images in all of these
lines (see Figs.~1, 2, 4a). In addition, we detected a large
number of new, weaker emission features, both compact and extended
diffuse, arc, and linear structures. We were able to resolve the
giant ring identified by Silich et~al.~(2002) into several
individual arc structures with different brightnesses and
different radii of curvature. Another faint outer arc can be
distinguished in the north outside the ring. All of the bright
emission features are observed on the background of weak diffuse
ionized-gas emission.

Our H$\alpha$-luminosity estimate for the galaxy is considerably lower than the value
obtained by Silich et~al.~(2002) and agrees well with its previous estimates by Lynds
et~al.~(1998) and Martin~(1998).

According to our long-slit spectroscopy presented above and our
preliminary data (Lozinskaya et~al.~2007), the
([NII]$\lambda6548$, 6583)/H$\alpha$
 intensity ratio in
VII~Zw~403 is $\sim0.03\mbox{--}0.035$; a similar value of 0.03 was obtained by
Martin~(1997). Accordingly, the (H$\alpha$~$+$~[NII]) luminosity determined from a
deeper image with a broad-band filter (see Table~3) gives
 $L(\textrm{H}\alpha) = 1.86\times 10^{39}$~erg~s$^{-1}$.

Using the formula for estimating the star formation rate (SFR)
from the H$\alpha$~luminosity, SFR $(M_{\odot}\,\mbox{yr}^{-1}) =
7.9\times 10^{-42} L(H\alpha)$~(Kennicutt~1998), we obtain SFR $=
0.012-0.015 M_{\odot}\mbox{yr}^{-1}$ from the derived luminosity
$L(H\alpha) = (1.49-1.86)\times 10^{39}$ erg~s$^{-1}$ for the
mass range~1 to $100 M_\odot$.

Based on our interferometric H$\alpha$ observations with FPI, we
constructed the line radial velocity and FWHM distributions in
the central region of the galaxy with an angular resolution of
$2\farcs2$. The H$\alpha$ FWHM in the star-forming region lies
within the range 60-70 to $100-120\km$. The ionized-gas velocity
in the star-forming region from $-120$ to $-90\km$ agrees well
with the neutral-gas velocity in the two dense HI~clouds
coincident with it that was measured by Thuan et~al.~(2004).

We failed to detect the expansion of the bright spherically
symmetric HII~regions with a velocity of $50-70\km$ that was
reported by Lynds et~al.~(1998). In our opinion, the most likely
cause of the difference between the results of Lynds
et~al.~(1998) and our results is that these authors used a slit
with a width of $2''$, which is clearly slightly larger than the
characteristic seeing at the KPNO Observatory. For a relatively
wide slit, the coherent structures resembling the velocity
ellipses of rapidly expanding shells in the spectra by Lynds
et~al. can result from brightness variations in the object across
the slit, so that the spatial and spectral features are mixed at
the output from the spectrograph. Unfortunately, the above paper
contains no seeing data and we cannot test this assumption. Note
also that the H$\alpha$ observations by Thuan et~al.~(1987) with
FPI and the observations by Martin~(1998) with an echelle
spectrograph with a spectral resolution corresponding to $11\km$
did not reveal any gas motions with such velocities anywhere in
the galaxy either.

We estimated the expansion velocity for shell~N4 $6.5-8''$
(140-180~pc) in size from the observed line broadening toward
the center to be $15-20\km$.

The corresponding kinematic age of the shell in the standard model by McCray and
Kafatos~(1987) is $\sim3$--4~Myr. The Ž'~association no.~5 responsible for the formation
of this shell contains 28 blue main-sequence Ž~stars and supergiants; the age of the
association is 5--6~Myr (Lynds et~al.~1998).

We estimated the upper limit for the expansion velocity of the
remaining bright HII~regions from the H$\alpha$, H$\beta$, and
[OIII] FWHMs to be $\textrm{V}_\textrm{exp} \leq 15-20\km$. The
corresponding kinematic age of the shells reaches 3--4~Myr or
more. This is considerably closer to the age of the compact
Ž'~associations responsible for their formation than the
kinematic age of the shells, $\sim1$~Myr, found by Lynds
et~al.~(1998) using the same standard model at the high expansion
velocity.

The age of the richest association no.~1 associated with shell~N1 was estimated by Lynds
et~al.~(1998) to be 4--5~Myr. Associations nos.~2, 3, 4, and~6 are considerably poorer
than associations nos.~1 and~5 and are also identified with the bright compact
HII~regions: associations nos.~3, 4, and~6 are identified with the HII~regions N6, N3,
and N5, respectively.

In addition to these relatively compact associations, HST observations also revealed
luminous blue stars in the entire central field of the galaxy. A more sparse grouping of
stars designated as ``center'' in Fig.~2 and Fig.~11 from Lynds et~al.~(1998) was
identified at the center of the star-forming region; the fainter diffuse shell N2
200--240~pc in size is associated with it. This mixed-age association contains a group
of blue supergiants ($\sim5$~Myr old) coincident with an older population (no younger
than 10~Myr), including 20 red supergiants of this age. Two populations of stars with
ages of 5 and 10~Myr or slightly older are also observed in the field designated as
``north'' and ``south'' by Lynds et~al.~(1998).

Thus, the central part of the galaxy reveals two starbursts~5 and
10~Myr ago; the interstellar medium shows traces of these two
bursts of the most recent star formation episode. All of the
mentioned bright HII~regions were identified with the associations
formed $\sim5$~Myr ago. The fainter filamentary and diffuse
H$\alpha$ emission regions distinguished in the entire central
region as well as the giant ring of ionized gas and the outer arc
outside it can most likely be associated with
 the older stellar population whose age reaches 10~Myr.

In general, the detected line broadening in the weak diffuse emission of ionized gas
outside the bright HII~regions may result from the superposition of several faint
background regions of ionized gas along the line of sight located in the gaseous galactic
disk and emitting at different radial velocities. However, another possibility is of
greater interest: the observed line broadening may be attributable to a more efficient
acceleration of the low-density gas by shock waves. The latter assumption can be tested
observationally based on the line intensity ratios in the spectrum. As we noted above,
the emission in the [OIII] and [SII] lines is actually enhanced with
respect to H$\alpha$ in the region of weak diffuse emission between the bright
HII~regions N4 and~N2 as well as to the north of regions~N3 and~N2 (see Fig.~2c). The
regions of maximum H$\alpha$ FWHM are also localized approximately here. This part of
the star-forming complex coincides with the region of comparatively weak 21-cm line
emission between the two densest HI~clouds, which also has a higher neutral-gas velocity
dispersion than that in the clouds (Thuan et~al. 2004). All of these facts may provide
evidence for a significant influence of the shock waves generated by stellar wind and
supernovae of the older population mentioned above on the low-density gas. The age of
10~Myr suggests that it is in these regions, not in the younger compact associations
associated with the bright shells, that the massive stars have already begun to go away
from the main sequence and supernova explosions occur in the star-forming region.

The question about the nature of the weak diffuse emission of
ionized gas in irregular galaxies has remained open for several
decades. The action of shock waves is actually considered as an
additional excitation source of forbidden lines. In VII~Zw~403,
the problem is compounded by the fact that the diffuse emission
of ionized gas is anomalously weak in this galaxy. (For this
reason, having observed 14 nearest irregular galaxies,
Martin~(1987, 1988) excluded VII~Zw 403 from her analysis  when
discussing the nature of the diffuse ionized gas.) Given the low
intensity of the H$\alpha$ emission in the two diffuse regions
mentioned above, our estimates of the line intensity ratios must
be confirmed by MPFS observations. We have already performed MPFS
observations; the emission spectrum of the diffuse ionized gas
and bright HII~regions in the galaxy will be analyzed in detail
in our forthcoming paper (Lozinskaya et~al.~2007). Here, we only
point out a preliminary result: our MPFS observations are
consistent with the filter observations presented here.

The very weak high-velocity H$\alpha$ and [OIII] wings detected
in the brightest supershell~N1 all the more require confirmation.
Such velocities, up to $-200- -300\km$ from the [OIII] line
center (at $\sim5\%$ of $I_{\textrm{max}}$) and the weak broad
H$\alpha$ pedestal corresponding to turbulent velocities with
FWHM up to $630\km$ (i.e., a gas velocity dispersion up to
$270\km$) have been detected in the galaxy VII~Zw~403 for the
first time. If these results were confirmed by independent
observations, such high velocities would unambiguously prove that
the gas accelerates at the shock front.

Note that the same two-component kinematics, faint high-velocity
features in slowly expanding shells, was observed in several
galactic supershells of similar sizes and was explained by the
action of a Wolf--Rayet stellar wind and/or by supernovae
explosions inside old supershells (see Lozinskaya (1998) and
references therein). Weak high-velocity H$\alpha$ wings are also
observed in bright HII~regions of other spiral galaxies (see
Rela\~{n}o and Beckman (2005) and references therein). However, in
both cases, the velocities are no larger than $100-200\km$ from
the line center.

Of greater interest is the detection of a weak broad component in
the H$\alpha$ line with FWHM=$1000\km$ in the central region of
the BCD~galaxy SBS~0335--052 (Pramskij and Moiseev 2003). The
relative intensity of the broad component accounted for $\sim2\%$
of the narrow component, i.e., approximately the same as that in
our case. Given the lower gas density in dwarf galaxies, the same
energy input from supernovae or stellar wind must actually
produce shock waves propagating with velocities higher than those
in spiral galaxies. As an example, we may note a recent paper by
Pustilnik et~al.~(2004) devoted to the BCD~galaxy HS~0837$+$4717,
who detected a broad (FWHM=$1300\km$) and higher-contrast (up to
$16\%$) pedestal in the H$\alpha$ line. The high-velocity
component has an unusually large Balmer decrement, which the
authors attribute both to electron impact line excitation and to
high internal extinction in the formation region of this
component.

Another problem that requires further studies is a detailed comparison of the ionized-
and neutral-gas kinematics in the central region of VII~Zw~403. It seems important for
establishing the genetic connection between the star-forming region and the dense
HI~complex at the center of the galaxy.

\section{Conclusions}

Thus, our long-slit spectroscopy with the same spectral resolution
as the spectral observations by Lynds et~al.~(1998), and our FPI
observations with three times the resolution of Lynds et al.
(1998) revealed no evidence of the expansion of the bright shells
with velocities of $50-70\km$.

The expansion velocity of the most extended shell~N4 determined
from the H$\alpha$ line broadening in the central region reaches
$15-20\km$. The upper limit for the expansion velocity of the
remaining bright HII regions estimated from the H$\alpha$,
H$\beta$, and [OIII] line FWHMs does not exceed $15-20\km$. The
corresponding kinematic age of the bright HII shells is 3--4~Myr
or more, in close agreement with the age of 4--5 Myr for the
compact OB~associations associated with them, as estimated by
Lynds et~al.~(1998).

Weak high-velocity wings of the [OIII] line (up to
$-200$...$-300\km$ from the line center) and H$\alpha$ line (with
FWHM as large as $600\km$ and a total flux in the broad component
reaching $4\%$ of the total intensity) were detected in the
brightest shell~N1 around the richest and youngest association
no.~1. Such velocities have been observed in the galaxy VII~Zw~403
for the first time.

We associate the faint filamentary and diffuse emission regions
that were identified  in nearly the entire central region of the
galaxy and the giant ring of ionized gas with the older and more
``sparse '' stellar population of the most recent starburst whose
age was estimated by Lynds et~al.~(1998) to be 10~Myr.

Our estimate of the total H$\alpha$ luminosity for the galaxy
$L_{\mbox{H}\alpha}=(1.49-1.86)\times10^{39}$~erg~s$^{-1}$ does
not confirm the measurements by Silich et~al.~(2002), but agrees
well with the earlier measurements by Lynds et~al.~(1998) and
Martin~(1998).

\begin{acknowledgements}
This work was supported by the Russian Foundation for Basic Research (project
nos.~04-02-16042 and 05-02-16454). A.V.~Moiseev thanks the Foundation for Support of
Russian Science for partial support. The work is based on the observational data
obtained with the 6-m SAO telescope financed by the Ministry of Science of Russia
(registration no. 01-43) and, in part, on the NASA/ESA Hubble Space Telescope data
retrieved from the Space Telescope Science Institute archive operated by the Association
of Universities for research in astronomy under contract with NASA (NAS~5-26555). We
also wish to thank N.~Podorvanyuk and M.~Yakim for help with the observations on the 6-m
SAO telescope.
\end{acknowledgements}

\textit{Translated by V.~Astakhov}


\begin{thebibliography}{}

\bibitem{}
Afanasiev V.L. and Moiseev A.V., 2005, Astron. Lett. \textbf{31}, 194 [astro-ph/0502095]

\bibitem{}%[3]%
Bomans D., 2001, Rev. Mod. Astron. \textbf{14}, 297

\bibitem{}%[4]%
Burstein D. and Heiles C., 1984, Astrophys. J.,~Suppl. Ser. \textbf{54}, 33

\bibitem{}%[5]%
Kennicutt R.C., 1998. Ann. Rev. Astron. Astrophys. \textbf{36}, 189

\bibitem{}%[7]%
Lira P., Lawrence A., and Johnson R.A., 2000, Mon. Not. R. Astron. Soc. \textbf{319}, 17


\bibitem{}%[10]%
Loose H.-H. and Thuan T.X., 1985,  \emph{Star-Forming Dwarf
Galaxies and Related Objects,} Ed. by D.~Kunth, T.X.~Thuan
(Frontieres, J. Tran Thanh Van Gif-sur-Yvette), 73

\bibitem{}%[8]%
Lozinskaya T.A., 1998. Astron. Lett. \textbf{24}, 237

\bibitem{}%[9]%
Lozinskaya T.A., Moiseev A.V., Avdeev V.Yu., and Egorov O.V.,
2007, in preparation

\bibitem{}%[6]%
Lynds R., Tolstoy E., O'Neil E.J., and Hunter D.A., Astron. J. \textbf{116}, 146

\bibitem{}%[12]%
Martin C.L., 1997, Astrophys. J. \textbf{491}, 561

\bibitem{}%[13]%
Martin C.L., 1998, Astrophys. J. \textbf{506}, 222

\bibitem{}%[11]%
McCray R. and Kafatos M., 1987, Astrophys. J. \textbf{317}, 190

\bibitem{}%[14]%
Moiseev A.V., 2002, Bull. Spec. Aastrophys. Obs. \textbf{54}, 74 [astro-ph/0211104]

\bibitem{}%[15]%
Ott J., Walter F., and Brinks E., 2005a, Mon. Not. R. Astron. Soc. \textbf{358}, 1423

\bibitem{}%[16]%
Ott J., Walter F., and Brinks E., 2005, Mon. Not. R. Astron. Soc. \textbf{358}, 1453

\bibitem{}%[17]%
Papaderos P., Fricke K.J., Thuan T.X., and Loose H.-H., 1994, Astron. Astrophys.
\textbf{291}, L13

\bibitem{}%[18]%
Pramskij A.G. and Moiseev A.V., 2003, Preprint N188, SAO RAN (Special Astrophysical
Observatory, Russian Academy of Science),  1.

\bibitem{}%[19]%
Pustilnik S., Kniazev A., Pramskij A., et al., 2004, Astron. Astrophys. \textbf{419}, 469


\bibitem{}%[20]%
Rela\~{n}o M., and Beckman J.E., Astron. Astrophys., 2005, \textbf{430}, 911

\bibitem{}%[25]%
Schulte-Ladbeck R.E., Hopp U., Greggio L., and Crone M.M., 1999a,
Astron. J. \textbf{118}, 2705


\bibitem{}%[24]%
Schulte-Ladbeck R.E., Hopp U., Crone M.M., and Greggio L., 1999b,
Astrophys. J. \textbf{525}, 709


\bibitem{}%[21]%
Silich S., Tenorio-Tagle G., Mu\~{n}oz-Tu\~{n}\'{o}n C., and Cair\'{o}s L.M., 2002, Astron. J. \textbf{123},
2438

\bibitem{}%[22]%
Thuan T.X., Hibbard J.E., and Levrier F., 2004, Astron. J. \textbf{128}, 717

\bibitem{}%[23]%
Thuan T.X., Williams T.B., and Malumuth E., 1987, in \emph{Starburst and Galaxy Evolution.
Procedings of the 22-th Moriond Astrophysical Meeting} (Frontieres, Gif-sur-Yvette,
1987), 151.
\end{thebibliography}
\end{document}